\documentclass[epj]{webofc}
\usepackage[varg]{txfonts} 
\usepackage{hyperref}

\newcommand*{\tabref}[1]{Table~\ref{tbl:#1}}

\renewcommand*{\eqref}[1]{Eq.~(\ref{eq:#1})}

\newcommand*{\figref}[1]{Fig.~(\ref{fig:#1})}

\wocname{ARENA-2016}
\woctitle{ARENA-2016}

\newcommand{\Emin}{\ensuremath{\mathcal{E}_{\rm min}}}
\newcommand{\Emax}{\ensuremath{\mathcal{E}_{\rm max}}}

\begin{document}
\title{Sensitivity of lunar particle-detection experiments}

\author{
 \firstname{Justin D.} \lastname{Bray}\inst{1}\fnsep\thanks{\email{justin.bray@manchester.ac.uk}}
}

\institute{
 JBCA, School of Physics \& Astronomy, University of Manchester, Manchester M13 9PL, UK
}

\abstract{
 The use of the Moon as a detector volume for ultra-high-energy neutrinos and cosmic rays, by searching for the Askaryan radio pulse produced when they interact in the lunar regolith, has been attempted by a range of projects over the past two decades.  In this contribution, I discuss some of the technical considerations relevant to these experiments, and their consequent sensitivity to ultra-high-energy particles.  I also discuss some possible future experiments, and highlight their potential.
}

\maketitle

\section{Introduction}
\label{sec:intro}


The detection of ultra-high-energy neutrinos and cosmic rays requires detectors with extremely large apertures.  One technique for obtaining such apertures is to remotely monitor solar-system bodies for radio pulses from the particle cascades produced when these particles interact, using these bodies themselves as detectors~\cite{gorham2004b}.  This technique has been employed in a commensal experiment with the Earth-orbiting FORTE satellite, with a non-detection setting a limit on the neutrino flux~\cite{lehtinen2004}; the proposed dedicated SWORD satellite~\cite{romero-wolf2013} may improve on this~\cite{motloch2014}.  Jupiter has also been proposed as a target~\cite{rimmer2014}, although its prospects appear poor~\cite{bray2016b}.

However, a target proposed before the above work~\cite{dagkesamanskii1989}, and the focus of most effort to date, is the Moon.  A particle cascade developing in the dense medium of the lunar regolith produces a strong Askaryan radio pulse~\cite{askaryan1962}, and the lunar Askaryan technique employs an Earth-based radio telescope --- or, potentially, a satellite in lunar orbit~\cite{ryabov2016} --- to search for this.  Eight such experiments have been published to date~\cite{hankins1996,gorham2004a,beresnyak2005,james2010,buitink2010,jaeger2010,spencer2010,bray2014a}; see Ref.~\cite{bray2016a} for a summary and analysis.  Below, I describe some of the limiting technical considerations for these experiments, and discuss their sensitivity to ultra-high-energy particles.

\section{Anticoincidence rejection}
\label{sec:anticoincidence}

A major difficulty for lunar Askaryan experiments is distinguishing between lunar-origin pulses and anthropogenic radio-frequency interference.  One solution is to use a telescope with multiple beams directed at different parts of the moon, and to apply an anticoincidence cut, rejecting any pulse that is detected in more than one beam.  A lunar-origin pulse should be detected only in the beam directed at the correct part of the moon, whereas local radio-frequency interference will generally be seen simultaneously in multiple beams.  Two recent experiments that used this approach are NuMoon~\cite{buitink2010} and LUNASKA~Parkes~\cite{bray2014a}.


However, telescope beams overlap to some degree, and a sufficiently strong lunar-origin pulse may be detected in multiple beams, and incorrectly classified as interference~\cite{bray2015a}.  This effectively sets an upper limit on the amplitude of an Askaryan pulse that can be detected by an experiment applying an anticoincidence cut.  Omitting this cut, however, leads to interference contaminating the sample of recorded events and raising the noise level, raising the lower limit on the amplitude of a detectable pulse.

As seen in \tabref{expvals}, the NuMoon experiment had only a narrow range between its upper and lower detection thresholds, leading to minimal sensitivity to ultra-high-energy particles; however, omitting the anticoincidence cut removes the upper limit while making little difference to the lower limit, restoring the sensitivity of the experiment.  The LUNASKA~Parkes experiment relied absolutely on its anticoincidence cut, having no appreciable sensitivity without it; but was sensitive to pulses across a broad range of amplitudes even with the anticoincidence cut included.  In both cases, the result was unintentional, with no thorough analysis of this issue beforehand.  For future experiments, a more sophisticated anticoincidence strategy will be developed, to minimise this issue~\cite{winchen_proc}.

\begin{table}
 \centering
 \caption{Minimum and maximum detectable radio pulse amplitudes for the NuMoon and LUNASKA~Parkes experiments, with and without anticoincidence cuts in their analyses~\cite{bray2016a}.  Note that these values are not a complete representation of the sensitivity of these experiments.}
 \label{tbl:expvals}
 \begin{tabular}{lccc}
  \hline
  Experiment     & Anticoincidence & \Emin  & \Emax \\
                 &                 & \multicolumn{2}{c}{($\mu$V/m/MHz)} \\
  \hline
  NuMoon         & yes             & 0.1365 & 0.1650 \\
                 & no              & 0.1453 & ---    \\
  LUNASKA~Parkes & yes             & 0.0053 & 0.0241 \\
                 & no              & $\infty$    & ---    \\
  \hline
 \end{tabular}
\end{table}

\section{Dedispersion}

An Askaryan pulse from the moon will be subjected to a frequency-dependent dispersive delay as it passes through the ionosphere, smearing it over a timescale longer than its original length.  This can be corrected by applying to the received signal a dedispersion filter that matches the dispersion characteristic of the ionosphere, defined by its slant total electron content (STEC).  Some past experiments with narrow bandwidths have simply neglected dispersion (e.g.~\cite{gorham2004a,jaeger2010}); others have corrected it in real time with a filter matched to the STEC (e.g.~\cite{james2010,bray2014a}), or recorded the received signal with its full time resolution, allowing the correction to be applied retrospectively (e.g.~\cite{buitink2010}).


Future lunar Askaryan experiments are likely to involve large bandwidths at low frequencies (\mbox{$\lesssim 500$}~MHz), for which dispersion is more severe.  In this case, the uncertainty in the STEC may be sufficient that, even after dedispersion, the pulse has some significant remnant dispersion characteristic; and the larger data rates of future experiments will make it prohibitive to record the received signal with its full time resolution, so retrospective processing is impractical.  The solution will likely involve a combination of parallel dedispersion at different dispersion characteristics, to cover the uncertainty range in STEC~\cite{james_proc,winchen_proc}; and more precise measurements of STEC, either through the direct use of line-of-sight measurements of STEC to satellites of the global positioning system (GPS), or through measurements of the STEC-dependent Faraday rotation of the lunar thermal emission~\cite{mcfadden2011}.

\section{Particle sensitivity}

I estimate the limits or potential limits set by some past and near-future lunar Askaryan experiments using an analytic model for the particle aperture~\cite{gayley2009,jeong2012,bray2016a}.  To reflect the effects of small-scale surface roughness, I use the only extant model~\cite{james2010}, which represents the aperture as a linear combination of two extreme cases: one in which these effects are ignored, and one in which they are maximised, by representing the surface as a series of small-scale facets with uncorrelated slopes.  Following this model, I take the parameter \mbox{$r = 0.6$}, describing the proportional contribution of the first case, as a compromise based on simulations of multiple experiments~\cite{bray2015a}, and scale $N_{\rm S}$, the number of facets along the length of the shower, according to the observing frequency of each experiment.  For the low-frequency experiments NuMoon and LOFAR, \mbox{$N_{\rm S} \sim 1$}, and so no correction for small-scale surface roughness is applied.

\begin{figure*}
 \centering
 \includegraphics[width=0.49\linewidth]{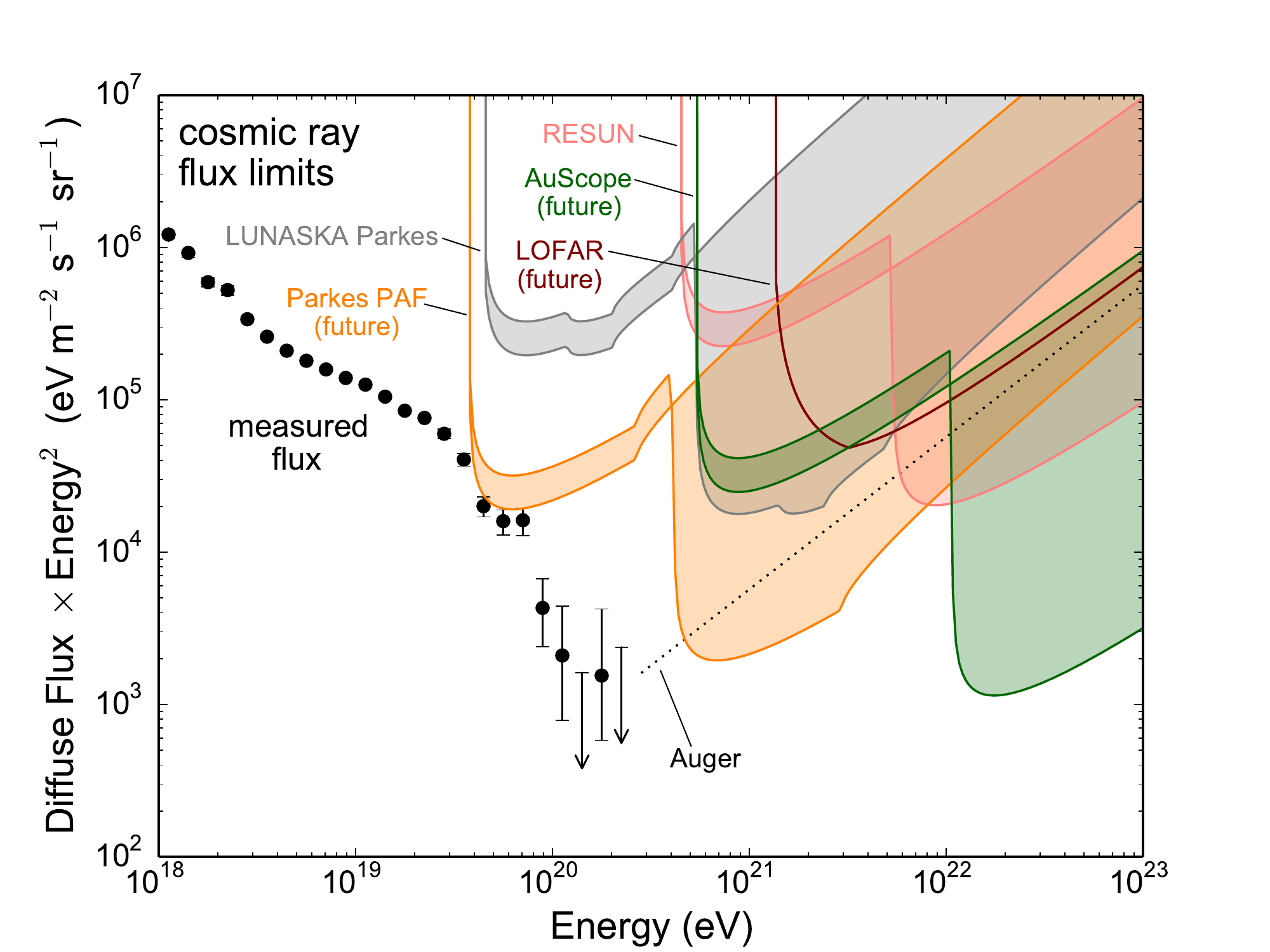}
 \includegraphics[width=0.49\linewidth]{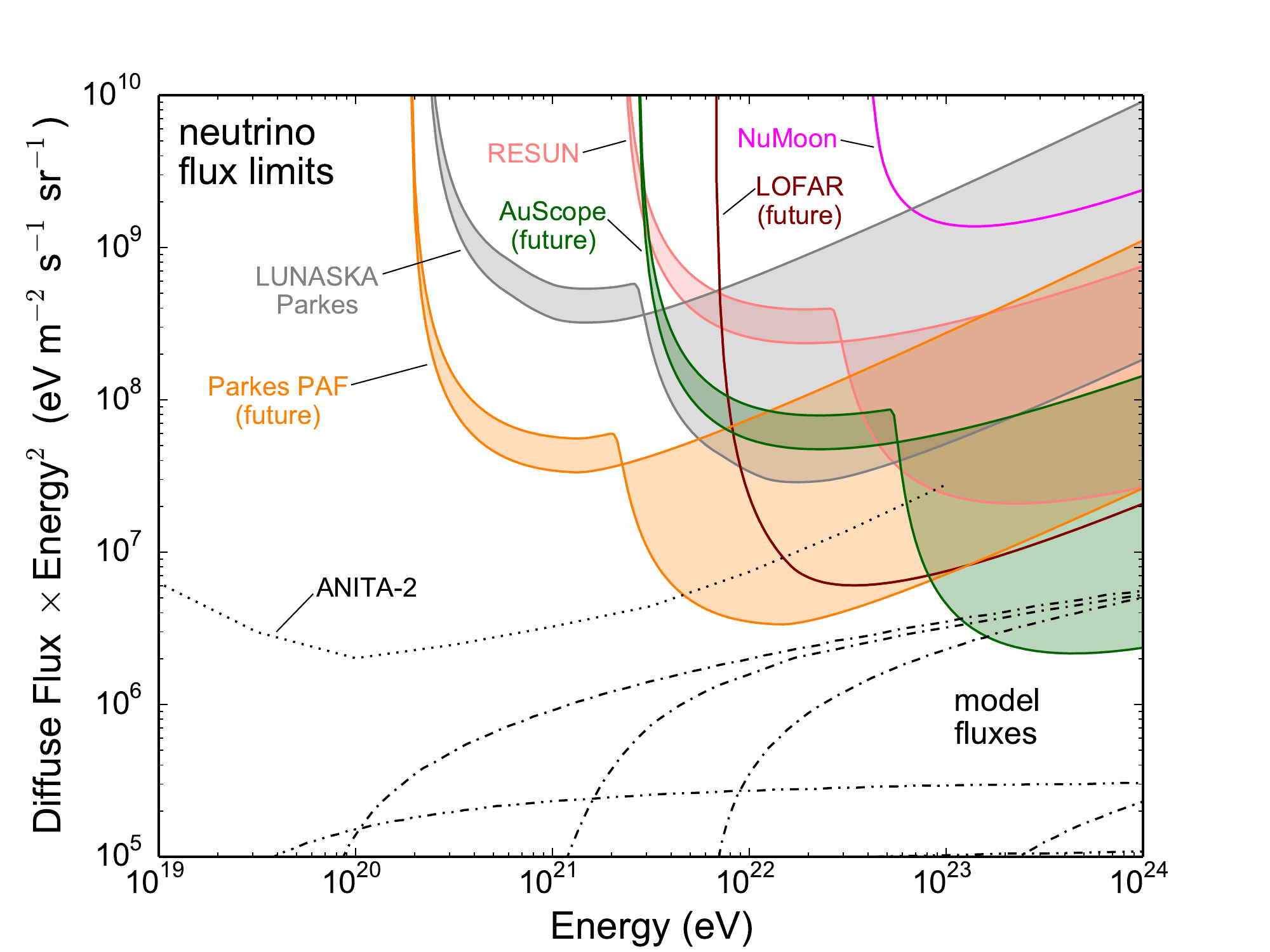}
 \caption{Limits on the fluxes of ultra-high-energy cosmic rays (left) and neutrinos (right) set by the NuMoon~\cite{buitink2010}, RESUN~\cite{jaeger2010} and LUNASKA Parkes~\cite{bray2014a} lunar Askaryan experiments, calculated as described in the text.  Also shown are potential limits that could be set by future lunar Askaryan experiments with nominal observing times of 200~h with LOFAR~\cite{singh2012,winchen_proc}, 200~h with a phased-array feed (PAF) on the Parkes radio telescope~\cite{bray2013}, or 2900~h with AuScope~\cite{bray2016a}.  Shading shows the range of uncertainty associated with models of the small-scale lunar surface roughness; the discontinuity is an artefact of the model.  Filled circles show the cosmic-ray flux measured by the Pierre Auger Observatory~\cite{abraham2010}, and dash-dotted lines show speculative models of the neutrino flux from exotic-physics models~\cite{berezinsky2011,lunardini2012}.  Dotted lines show limits established by the Pierre Auger Observatory~\cite{abraham2010} and the ANITA-2 experiment~\cite{gorham2010}.  All limits are 90\%-confidence model-independent differential limits~\cite{lehtinen2004}.}
 \label{fig:flux_all}
\end{figure*}

Results are shown in \figref{flux_all}.  Uncertainty associated with the effects of small-scale surface roughness strongly affects the estimated limits for high-frequency experiments.  Depending on these effects, near-future high-frequency experiments may or may not improve on existing neutrino limits at ultra-high energies.  A planned low-frequency experiment with LOFAR~\cite{singh2012,winchen_proc} will do so in any case, improving current constraints on neutrino fluxes predicted by exotic-physics models.  A proposed experiment with a phased-array feed on the Parkes radio telescope~\cite{bray2013}, or on a similar large single dish, could potentially achieve the first detection of cosmic rays with this technique.  A proposed experiment (not shown here) with the low-frequency component of the Square Kilometre Array~\cite{bray2014b,james_proc} would have a substantially greater sensitivity to both neutrinos and cosmic rays, but is a more distant prospect, as this telescope is planned for construction in 2018--2023.

\section{Conclusion}

The technical considerations for lunar Askaryan experiments are challenging, but are now well understood, and future experiments can be expected to address them correctly.  The major theoretical problem remaining in this field is modelling the effects of small-scale lunar surface roughness, which have a dramatic effect on the particle aperture for high-frequency experiments.

Experiments practical on a near-future timescale have the potential to substantially improve on the current state of this field.  A lunar Askaryan experiment with a phased-array feed on a large single dish could plausibly achieve the first detection with this technique of an ultra-high-energy cosmic ray.  The planned experiment with LOFAR will constrain neutrino fluxes predicted by exotic-physics models, as well as being a technical pathfinder for a future experiment with the Square Kilometre Array.

\begin{acknowledgement}
 I acknowledge support from ERC-StG 307215 (LODESTONE).
\end{acknowledgement}

\bibliography{all}

\end{document}